\pdfoutput=1
\documentclass[12pt,a4paper]{article}

\usepackage{ifthen} 
\newboolean{pdflatex}
\setboolean{pdflatex}{true} 

\newboolean{articletitles}
\setboolean{articletitles}{true} 

\newboolean{uprightparticles}
\setboolean{uprightparticles}{false} 

\newboolean{inbibliography}
\setboolean{inbibliography}{false} 

\newboolean{wordcount}
\setboolean{wordcount}{false} 

\usepackage{microtype}
\usepackage{lineno}  
\usepackage{xspace} 
\usepackage{caption} 

\usepackage{graphicx}  
\usepackage{color}
\usepackage{colortbl}
\graphicspath{{./figs/}} 

\usepackage{amsmath} 
\usepackage{amssymb}
\usepackage{amsfonts}
\usepackage{upgreek} 

\newcommand*\patchAmsMathEnvironmentForLineno[1]{%
\expandafter\let\csname old#1\expandafter\endcsname\csname #1\endcsname
\expandafter\let\csname oldend#1\expandafter\endcsname\csname
end#1\endcsname
 \renewenvironment{#1}%
   {\linenomath\csname old#1\endcsname}%
   {\csname oldend#1\endcsname\endlinenomath}%
}
\newcommand*\patchBothAmsMathEnvironmentsForLineno[1]{%
  \patchAmsMathEnvironmentForLineno{#1}%
  \patchAmsMathEnvironmentForLineno{#1*}%
}
\AtBeginDocument{%
\patchBothAmsMathEnvironmentsForLineno{equation}%
\patchBothAmsMathEnvironmentsForLineno{align}%
\patchBothAmsMathEnvironmentsForLineno{flalign}%
\patchBothAmsMathEnvironmentsForLineno{alignat}%
\patchBothAmsMathEnvironmentsForLineno{gather}%
\patchBothAmsMathEnvironmentsForLineno{multline}%
\patchBothAmsMathEnvironmentsForLineno{eqnarray}%
}

\usepackage{hyperref}    
\usepackage[all]{hypcap} 

\def\lhcb {\mbox{LHCb}\xspace}

\def\babar  {\mbox{BaBar}\xspace}

\def\dzero  {\mbox{D0}\xspace}

\def\MagUp {\mbox{\em Mag\kern -0.05em Up}\xspace}

\ifthenelse{\boolean{uprightparticles}}%
{

 \def\Pmu         {\ensuremath{\upmu}\xspace}                 
 \def\Pnu         {\ensuremath{\upnu}\xspace}                 
                  
 \def\Ppi         {\ensuremath{\uppi}\xspace}

 \def\Ptau        {\ensuremath{\uptau}\xspace}

 \def\Ppsi        {\ensuremath{\uppsi}\xspace}

 \def\PDelta      {\ensuremath{\Delta}\xspace}                 
 \def\PXi      {\ensuremath{\Xi}\xspace}                 
 \def\PLambda      {\ensuremath{\Lambda}\xspace}                 
 \def\PSigma      {\ensuremath{\Sigma}\xspace}                 
 \def\POmega      {\ensuremath{\Omega}\xspace}                 
 \def\PUpsilon      {\ensuremath{\Upsilon}\xspace}                 
 
 \def\PB      {\ensuremath{\mathrm{B}}\xspace}                 
                  
 \def\PD      {\ensuremath{\mathrm{D}}\xspace}

 \def\PJ      {\ensuremath{\mathrm{J}}\xspace}                 
 \def\PK      {\ensuremath{\mathrm{K}}\xspace}

 \def\PX      {\ensuremath{\mathrm{X}}\xspace}

 \def\Pb      {\ensuremath{\mathrm{b}}\xspace}                 
 \def\Pc      {\ensuremath{\mathrm{c}}\xspace}                 
 \def\Pd      {\ensuremath{\mathrm{d}}\xspace}

 \def\Pi      {\ensuremath{\mathrm{i}}\xspace}

 \def\Pp      {\ensuremath{\mathrm{p}}\xspace}

 \def\Ps      {\ensuremath{\mathrm{s}}\xspace}

}
{

 \def\Pmu         {\ensuremath{\mu}\xspace}                 
 \def\Pnu         {\ensuremath{\nu}\xspace}                 
                  
 \def\Ppi         {\ensuremath{\pi}\xspace}

 \def\Ptau        {\ensuremath{\tau}\xspace}

 \def\Ppsi        {\ensuremath{\psi}\xspace}                 
                  
 \mathchardef\PDelta="7101
 \mathchardef\PXi="7104
 \mathchardef\PLambda="7103
 \mathchardef\PSigma="7106
 \mathchardef\POmega="710A
 \mathchardef\PUpsilon="7107
                  
 \def\PB      {\ensuremath{B}\xspace}                 
                  
 \def\PD      {\ensuremath{D}\xspace}

 \def\PJ      {\ensuremath{J}\xspace}                 
 \def\PK      {\ensuremath{K}\xspace}

 \def\PX      {\ensuremath{X}\xspace}

 \def\Pb      {\ensuremath{b}\xspace}                 
 \def\Pc      {\ensuremath{c}\xspace}                 
 \def\Pd      {\ensuremath{d}\xspace}

 \def\Pi      {\ensuremath{i}\xspace}

 \def\Pp      {\ensuremath{p}\xspace}

 \def\Ps      {\ensuremath{s}\xspace}

}

\makeatletter
\ifcase \@ptsize \relax
  \newcommand{\miniscule}{\@setfontsize\miniscule{4}{5}}
\or
  \newcommand{\miniscule}{\@setfontsize\miniscule{5}{6}}
\or
  \newcommand{\miniscule}{\@setfontsize\miniscule{5}{6}}
\fi
\makeatother

\DeclareRobustCommand{\optbar}[1]{\shortstack{{\miniscule (\rule[.5ex]{0.8em}{.18mm})}
  \\ [-.7ex] $#1$}}

\def\mup        {{\ensuremath{\Pmu^+}}\xspace}
\def\mun        {{\ensuremath{\Pmu^-}}\xspace} 

\def\taup       {{\ensuremath{\Ptau^+}}\xspace}

\def\neu        {{\ensuremath{\Pnu}}\xspace}
\def\neub       {{\ensuremath{\overline{\Pnu}}}\xspace}
\def\neum       {{\ensuremath{\neu_\mu}}\xspace}
\def\neumb      {{\ensuremath{\neub_\mu}}\xspace}

\def\dquark    {{\ensuremath{\Pd}}\xspace}

\def\squark    {{\ensuremath{\Ps}}\xspace}

\def\cquark    {{\ensuremath{\Pc}}\xspace}

\def\bquark    {{\ensuremath{\Pb}}\xspace}

\def\pion   {{\ensuremath{\Ppi}}\xspace}

\def\pip    {{\ensuremath{\pion^+}}\xspace}
\def\pim    {{\ensuremath{\pion^-}}\xspace}

\def\kaon    {{\ensuremath{\PK}}\xspace}
  \def\Kbar    {{\kern 0.2em\overline{\kern -0.2em \PK}{}}\xspace}

\def\KorKbar    {\kern 0.18em\optbar{\kern -0.18em K}{}\xspace}
\def\Kz      {{\ensuremath{\kaon^0}}\xspace}

\def\Kp      {{\ensuremath{\kaon^+}}\xspace}
\def\Km      {{\ensuremath{\kaon^-}}\xspace}

  \def\Dbar    {{\kern 0.2em\overline{\kern -0.2em \PD}{}}\xspace}
\def\D       {{\ensuremath{\PD}}\xspace}

\def\DorDbar    {\kern 0.18em\optbar{\kern -0.18em D}{}\xspace}
\def\Dz      {{\ensuremath{\D^0}}\xspace}
\def\Dzb     {{\ensuremath{\Dbar{}^0}}\xspace}

\def\Dm      {{\ensuremath{\D^-}}\xspace}

\def\Dstarm  {{\ensuremath{\D^{*-}}}\xspace}

\def\B       {{\ensuremath{\PB}}\xspace}
\def\Bbar    {{\ensuremath{\kern 0.18em\overline{\kern -0.18em \PB}{}}}\xspace}

\def\BorBbar    {\kern 0.18em\optbar{\kern -0.18em B}{}\xspace}
\def\Bz      {{\ensuremath{\B^0}}\xspace}

\def\Bu      {{\ensuremath{\B^+}}\xspace}

\def\Bd      {{\ensuremath{\B^0}}\xspace}
\def\Bs      {{\ensuremath{\B^0_\squark}}\xspace}

\def\Bdb     {{\ensuremath{\Bbar{}^0}}\xspace}

\def\jpsi     {{\ensuremath{{\PJ\mskip -3mu/\mskip -2mu\Ppsi\mskip 2mu}}}\xspace}

  \def\Y#1S{\ensuremath{\PUpsilon{(#1S)}}\xspace}

\def\Lz          {{\ensuremath{\PLambda}}\xspace}
\def\Lbar        {{\ensuremath{\kern 0.1em\overline{\kern -0.1em\PLambda}}}\xspace}
\def\LorLbar    {\kern 0.18em\optbar{\kern -0.18em \PLambda}{}\xspace}

\def\Lb      {{\ensuremath{\Lz^0_\bquark}}\xspace}

\def\Lc      {{\ensuremath{\Lz^+_\cquark}}\xspace}

\newcommand{\decay}[2]{\ensuremath{#1\!\to #2}\xspace}         

\def\to                 {\ensuremath{\rightarrow}\xspace}

\def\CP                {{\ensuremath{C\!P}}\xspace}

\newcommand{\dmd}{{\ensuremath{\Delta m_{\dquark}}}\xspace}

\newcommand{\DGd}{{\ensuremath{\Delta\Gamma_{\dquark}}}\xspace}

\newcommand{\Gd}{{\ensuremath{\Gamma_{\dquark}}}\xspace}

\def\AT#1     {\ensuremath{A_{\mathrm{T}}^{#1}}\xspace}           

\def\C#1      {\ensuremath{\mathcal{C}_{#1}}\xspace}                       
\def\Cp#1     {\ensuremath{\mathcal{C}_{#1}^{'}}\xspace}                    
\def\Ceff#1   {\ensuremath{\mathcal{C}_{#1}^{\mathrm{(eff)}}}\xspace}        
\def\Cpeff#1  {\ensuremath{\mathcal{C}_{#1}^{'\mathrm{(eff)}}}\xspace}       
\def\Ope#1    {\ensuremath{\mathcal{O}_{#1}}\xspace}                       
\def\Opep#1   {\ensuremath{\mathcal{O}_{#1}^{'}}\xspace}                    

\newcommand{\tev}{\ensuremath{\mathrm{\,Te\kern -0.1em V}}\xspace}
\newcommand{\gev}{\ensuremath{\mathrm{\,Ge\kern -0.1em V}}\xspace}
\newcommand{\mev}{\ensuremath{\mathrm{\,Me\kern -0.1em V}}\xspace}
\newcommand{\kev}{\ensuremath{\mathrm{\,ke\kern -0.1em V}}\xspace}
\newcommand{\ev}{\ensuremath{\mathrm{\,e\kern -0.1em V}}\xspace}
\newcommand{\gevc}{\ensuremath{{\mathrm{\,Ge\kern -0.1em V\!/}c}}\xspace}
\newcommand{\mevc}{\ensuremath{{\mathrm{\,Me\kern -0.1em V\!/}c}}\xspace}
\newcommand{\gevcc}{\ensuremath{{\mathrm{\,Ge\kern -0.1em V\!/}c^2}}\xspace}
\newcommand{\gevgevcccc}{\ensuremath{{\mathrm{\,Ge\kern -0.1em V^2\!/}c^4}}\xspace}
\newcommand{\mevcc}{\ensuremath{{\mathrm{\,Me\kern -0.1em V\!/}c^2}}\xspace}

\def\mum  {\ensuremath{{\,\upmu\rm m}}\xspace}

\def\invfb   {\ensuremath{\mbox{\,fb}^{-1}}\xspace}

\def\ps   {\ensuremath{{\rm \,ps}}\xspace}

\def\gsim{{~\raise.15em\hbox{$>$}\kern-.85em
          \lower.35em\hbox{$\sim$}~}\xspace}
\def\lsim{{~\raise.15em\hbox{$<$}\kern-.85em
          \lower.35em\hbox{$\sim$}~}\xspace}

\newcommand{\mean}[1]{\ensuremath{\left\langle #1 \right\rangle}} 

\def\ptot       {\mbox{$p$}\xspace}
\def\pt         {\mbox{$p_{\rm T}$}\xspace}

\def\tell1  {TELL1\xspace}
\def\ukl1   {UKL1\xspace}

\usepackage{cite} 
\usepackage{mciteplus}

\def\myTitle  {Measurement of the semileptonic $C\!P$ asymmetry in $B^0$--$\kern 0.18em\overline{\kern -0.18em B}{}^0$ mixing}

\def\asldVal  {-0.02}
\def\asldStat {0.19}
\def\asldSyst {0.30}

\def\asldValDMu   {-0.19}
\def\asldStatDMu   {0.21}
\def\asldValDstMu  {0.77}
\def\asldStatDstMu {0.45}

\def\ApValSeven  {-0.66}
\def\ApStatSeven {0.26}
\def\ApSystSeven {0.22}
\def\ApValEight  {-0.48}
\def\ApStatEight {0.15}
\def\ApSystEight {0.17}

\def\fbar    {{\ensuremath{\kern 0.18em\overline{\kern -0.18em f}{}}}\xspace}
\def\forfbar    {\kern 0.18em\optbar{\kern -0.18em f}{}\xspace}

\def\asld   {\ensuremath{a_{\rm sl}^{d}}\xspace}
\def\asls   {\ensuremath{a_{\rm sl}^{s}}\xspace}

\def\AP     {\ensuremath{A_{\rm P}}\xspace}

\def\AD     {\ensuremath{A_{\rm D}}\xspace}

\def\BdToDstMuNu    {\decay{\Bd}{\Dstarm\mup\neum X}}
\def\BdToDmMuNu     {\decay{\Bd}{\Dm\mup\neum X}}
\def\DstarMode      {\BdToDstMuNu}

\def\DmMode         {\BdToDmMuNu}

\def\DmToKpipi      {\decay{\Dm}{\Kp\pim\pim}}

\def\DzbToKpi       {\decay{\Dzb}{\Kp\pim}}

\def\DstDm          {\ensuremath{\D^{(*)-}}\xspace}
\def\DstDp          {\ensuremath{\D^{(*)+}}\xspace}

\def\DMu            {\Dm\mup}
\def\DstMu          {\Dstarm\mup}
\def\DstDMu         {\DstDm\mup}

\def\DstDMuNu       {\DstDm\mup\neum}

\newcommand{\gevcNS}{\ensuremath{{\mathrm{Ge\kern -0.1em V\!/}c}}\xspace}

\DeclareGraphicsExtensions{.pdf,.PDF,png,.PNG}

\def\figOneWidth{0.65}
\def\figTwoWidth{0.7}
\def\breakEquation{}
\usepackage{afterpage}

\begin{document}

\renewcommand{\thefootnote}{\fnsymbol{footnote}}
\setcounter{footnote}{1}


\begin{titlepage}
\pagenumbering{roman}

\vspace*{-1.5cm}
\centerline{\large EUROPEAN ORGANIZATION FOR NUCLEAR RESEARCH (CERN)}
\vspace*{1.5cm}
\hspace*{-0.5cm}
\begin{tabular*}{\linewidth}{lc@{\extracolsep{\fill}}r}
\ifthenelse{\boolean{pdflatex}}
{\vspace*{-2.7cm}\mbox{\!\!\!\includegraphics[width=.14\textwidth]{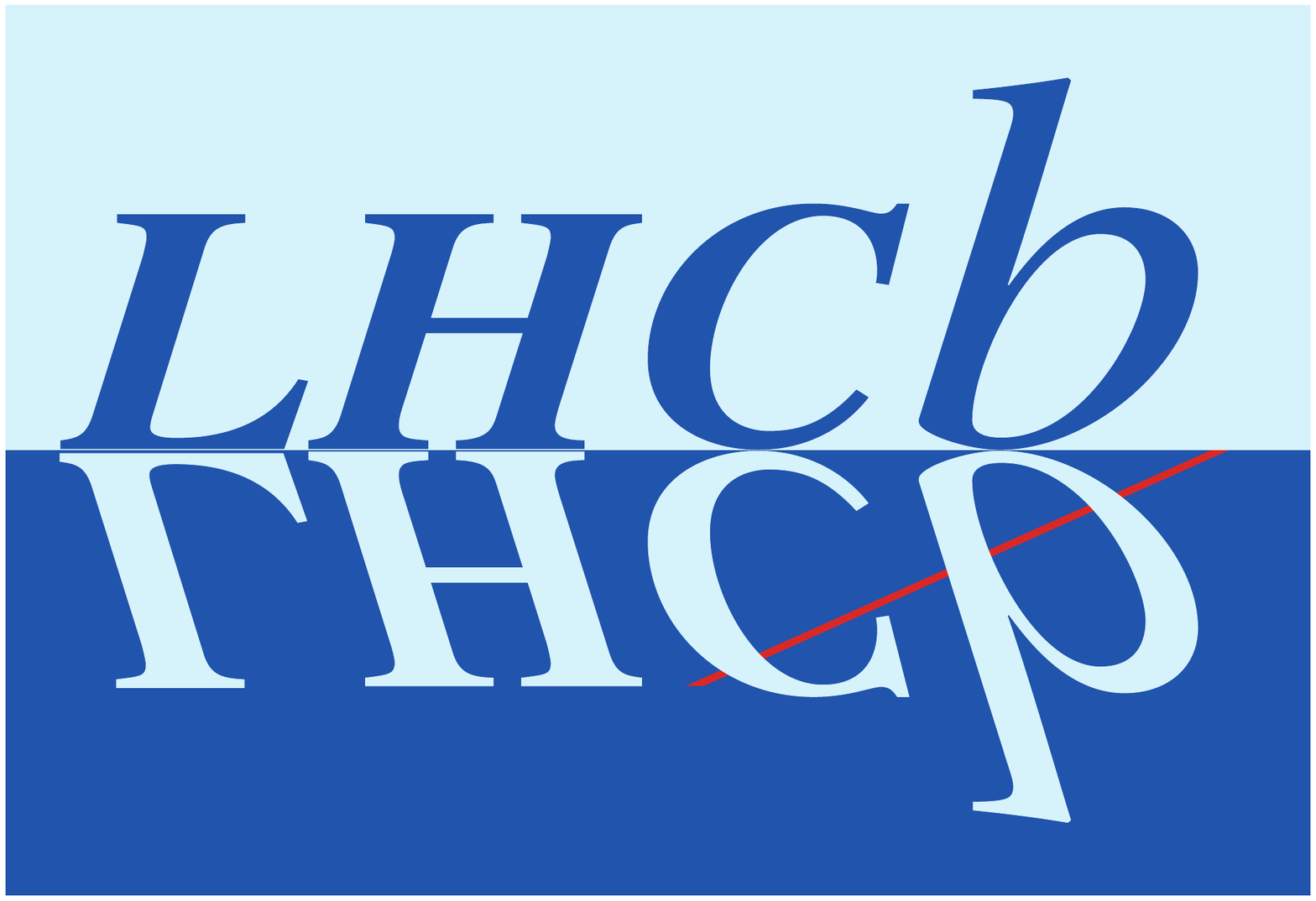}} & &}%
{\vspace*{-1.2cm}\mbox{\!\!\!\includegraphics[width=.12\textwidth]{lhcb-logo.eps}} & &}%
\\
 & & CERN-PH-EP-2014-237 \\  
 & & LHCb-PAPER-2014-053 \\  
 & & January 29, 2015 \\
 & & \\
\end{tabular*}

\vspace*{1.0cm}

{\bf\boldmath\huge
\begin{center}
\myTitle
\end{center}
}

\vspace*{1.0cm}

\begin{center}
The LHCb collaboration\footnote{Authors are listed at the end of this Letter.}
\end{center}

\vspace{\fill}

\begin{abstract}
  \noindent
The semileptonic $C\!P$ asymmetry in $B^0$--$\kern 0.18em\overline{\kern -0.18em
  B}{}^0$ mixing, $a_{\rm sl}^d$, is measured in proton--proton collision data,
corresponding to an integrated luminosity of 3.0~fb$^{-1}$, recorded by the LHCb
experiment. Semileptonic $B^0$ decays are reconstructed in the inclusive final
states $D^-\mu^+$ and $D^{*-}\mu^+$, where the $D^-$ meson decays into the
$K^+\pi^-\pi^-$ final state, and the $D^{*-}$ meson into the $\kern
0.18em\overline{\kern -0.18em D}{}^0(\rightarrow K^+\pi^-)\pi^-$ final
state. The asymmetry between the numbers of $D^{(*)-}\mu^+$ and $D^{(*)+}\mu^-$
decays is measured as a function of the decay time of the $B^0$ mesons. The
$C\!P$ asymmetry is measured to be \mbox{$a_{\rm sl}^d = (\asldVal \pm \asldStat \pm
\asldSyst)\%$}, where the first uncertainty is statistical and the second
systematic. This is the most precise measurement of $a_{\rm sl}^d$ to date and
is consistent with the prediction from the Standard Model.

\end{abstract}

\vspace*{1.0cm}

\begin{center}
  Published in Phys.~Rev.~Lett.
\end{center}

\vspace{\fill}

{\footnotesize 
\centerline{\copyright~CERN on behalf of the \lhcb collaboration, license \href{http://creativecommons.org/licenses/by/4.0/}{CC-BY-4.0}.}}
\vspace*{2mm}

\end{titlepage}


\newpage
\setcounter{page}{2}
\mbox{~}
%
%
%
%

\cleardoublepage


\renewcommand{\thefootnote}{\arabic{footnote}}
\setcounter{footnote}{0}


\pagestyle{plain} 
\setcounter{page}{1}
\pagenumbering{arabic}


\noindent
The inclusive charge asymmetry measured by the \dzero collaboration in events
with same charge dimuons~\cite{Abazov:2013uma} shows one of the largest
discrepancies with the Standard Model and it may be a first hint of physics
beyond our current understanding  (e.g., Refs.~\cite{Dobrescu:2010rh, DescotesGenon:2012kr, Sahoo:2013vuf}). This asymmetry is sensitive to \CP violation
in the mixing of neutral \B mesons.
The neutral \Bd meson and its antiparticle \Bdb are flavour eigenstates, formed
from a mixture of two mass eigenstates. The time evolution of this two-state
system results in flavour-changing $\Bd \to \Bdb$ and $\Bdb \to \Bd$
transitions. Violation of charge-parity (\CP) symmetry may occur due to this
process if the probability for a \Bd meson to transform into a \Bdb meson is
different from the reverse process. When a meson produced in the \Bd eigenstate
decays semileptonically to a final state $f$, the charge of the lepton reveals
the meson flavour at the time of decay. In such decays, ``wrong-sign''
transitions, like $\Bd \to \fbar$, can only happen due to the transition $\Bd
\to \Bdb \to \fbar$. The flavour-specific (semileptonic) asymmetry is defined in
terms of partial decay rates $\Gamma$ as
\begin{align}
 \asld \equiv \frac{\Gamma(\Bdb \to f) - \Gamma(\Bd \to \fbar)}
       {\Gamma(\Bdb \to f) + \Gamma(\Bd \to \fbar)} 
       \approx \frac{\DGd}{\dmd}\tan\phi_d^{12} \ ,
\label{eq:asl}
\end{align}
and is expressed in terms of the difference between the masses (\dmd) and widths
(\DGd) of the mass eigenstates, and the \CP-violating phase
$\phi_d^{12}$~\cite{Lenz:2006hd}. The Standard Model (SM) prediction,
$\asld=(-4.1\pm0.6) \times 10^{-4}$~\cite{Lenz:2011ti}, is small compared to
experimental sensitivities. However, \asld may be enhanced by virtual
contributions from particles that exist in extensions to the
SM~\cite{Lenz:2012az}.

The current most precise measurements are $\asld = (0.06 \pm
0.17^{+0.38}_{-0.32})\%$ by the \babar collaboration~\cite{Lees:2013sua} and
$\asld=(0.68\pm0.45\pm0.14)\%$ by the \dzero
collaboration~\cite{Abazov:2012hha}, where the first uncertainties are
statistical and the second systematic. The \dzero dimuon
asymmetry, which is related to a linear combination of the
semileptonic asymmetries in the \Bd and \Bs systems, disagrees with the
theoretical predictions by 3.6 standard deviations. The \lhcb collaboration has
previously measured the semileptonic \CP asymmetry in the \Bs system,
\asls~\cite{LHCb-PAPER-2013-033}, consistent with the SM. Improved experimental
constraints are also required on \asld to confirm or falsify the \dzero anomaly.

In this analysis, \asld is measured using semileptonic \DmMode and \DstarMode
decays, where \PX denotes any additional particles due to possible feed-down
from \taup decays into $\mup\PX$ and higher-resonance \D decays into
$\DstDm\PX$. The inclusion of charge-conjugate processes is implied. The signal
is reconstructed from \DstDMu pairs, with the charm mesons reconstructed from
\DmToKpipi and $\Dstarm\to\Dzb(\to\Kp\pim)\pim$ decays. A measurement of \asld
using the quantities in Eq.~(\ref{eq:asl}) requires determining (tagging) the
flavour of the \Bd meson at production. Since this is inefficient in hadron
collisions, \asld is instead determined from the untagged decay rates. The
number of observed final states as a function of the \Bd decay time is expressed
as
\begin{align}  
   N(t) \propto e^{-\Gd t}
   \biggl[ & 1 + \zeta \AD + \zeta \frac{\asld}{2} - \breakEquation
            \zeta \left(\AP + \frac{\asld}{2} \right) \cos\dmd\,t \biggr] \ ,
\label{eq:decayrates}
\end{align}
where \Gd is the \Bd decay width, and $\zeta=+1(-1)$ for the $f$ ($\fbar$) final
state. The asymmetry due to differences in detection efficiencies,
$\varepsilon$, between $f$ and \fbar final states, \mbox{$\AD \equiv
[\varepsilon(f) - \varepsilon(\fbar)] / [\varepsilon(f) + \varepsilon(\fbar)]$},
is determined using control samples of data, as described later. The asymmetry
in the \Bdb and \Bd effective production cross sections, \mbox{$\AP \equiv
[ \sigma(\Bdb)-\sigma(\Bd) ] / [ \sigma(\Bdb)+\sigma(\Bd) ]$}, and
\asld are determined simultaneously in a fit to the time-dependent rate of
Eq.~(\ref{eq:decayrates}). Effects from higher-order asymmetry terms and a
non-zero \DGd, taken from experimental bounds~\cite{PDG2014}, result in biases
of less than $10^{-4}$ on \asld and are ignored. The amount of direct \CP
violation in the Cabibbo-favoured decays \DmToKpipi and \DzbToKpi is assumed to
be negligible. The observed decay time of the semileptonic signal candidates is
corrected using simulation since the final state is only partially
reconstructed.

The \lhcb detector~\cite{Alves:2008zz} includes a high-precision tracking system
with a dipole magnet, providing a measurement of momentum (\ptot) and impact
parameter (IP) for charged particles. The IP, defined as the minimum distance of
a track to a proton--proton ($pp$) interaction vertex, is measured with a
precision of about $20\mum$ for high-momentum tracks. The polarity of the
magnetic field is regularly reversed during data taking. Particle identification
(PID) is provided by ring-imaging Cherenkov detectors, a calorimeter and a muon
system.  The trigger~\cite{LHCb-DP-2012-004} consists of a hardware stage, based
on information from the calorimeter and muon systems, followed by a software
stage, which applies a full event reconstruction.

In the simulation, $pp$ collisions are generated~\cite{Sjostrand:2006za,
  *Sjostrand:2007gs, *LHCb-PROC-2010-056, *Lange:2001uf, *Golonka:2005pn}, and
the interactions of the outgoing particles with the detector are
modelled~\cite{Allison:2006ve, *Agostinelli:2002hh, *LHCb-PROC-2011-006}.  The
\B mesons are required to decay semileptonically to a muon, a neutrino and a
\DstDm meson. Feed-down from higher \D resonances and \Ptau decays is based on
branching fractions, either measured~\cite{PDG2014} or estimated assuming
isospin symmetry.

The data used in this analysis correspond to a luminosity of 3.0\invfb, of which
1.0 (2.0)\invfb was taken in 2011 (2012) at a $pp$ centre-of-mass energy of 7
(8)\tev.  The selection of candidates relies on the signatures of high-momentum
tracks and displaced vertices from the \Bd, \Dm and \Dzb decays. Candidate
events are first required to pass the hardware trigger, which selects muons with
momentum transverse to the beam direction (\pt) larger than $1.64~(1.76)\gevc$
for the 2011 (2012) data. In a first stage of the software trigger, the muon
is required to have a large IP. In a second stage, the muon and at least one
of the \DstDm decay products are required to be consistent with the topological
signature of \bquark-hadron decays~\cite{LHCb-DP-2012-004}.

To suppress background, it is required that the tracks from the \Bd candidates
do not point back to any $pp$ interaction vertex. The muon, kaon and pion
candidates are required to be well identified by the PID system. Tracks from the
\Dm, \Dzb and \Bd candidates are required to form well-defined vertices. For the
\DstMu final state, the difference between the \Dstarm and \Dzb masses should be
between 144 and $147\mevcc$. The mass of the \DstDMu final state is required to
be between $3.0$ and $5.2\gevcc$ to allow for missing particles in the
final state; the upper limit removes background from four-body \bquark-hadron
decays. Misreconstructed \D candidates made from random combinations of tracks
are suppressed by requiring that the \Dm or \Dzb decay time is larger than
$0.1\ps$. The contribution from charm decays directly produced in the $pp$
interaction (prompt \D) is reduced to below $0.1\%$ by requiring \Dm and \Dzb
candidates to have an IP larger than $50\mum$.

Detection asymmetries caused by left-right asymmetries in the reconstruction
efficiency change sign when the polarity of the LHCb magnet is inverted.  Other
asymmetries, such as those induced by differing nuclear cross sections for \Kp
and \Km mesons, do not depend on the magnet polarity. The detection asymmetry
of the $\Kp\pim\pim\mup$ final state is factorized into a $\pim\mup$ component,
where the pion is the hard one (i.e., from the \Dzb decay or the higher-\pt pion
in the \Dm decay), and a $\Kp\pim$ component, where the pion is the soft one.

For the $\pim\mup$ component, any asymmetry arising from the different tracking
efficiencies is suppressed by weighting the signal candidates such that the muon
and hard pion have the same \pt and pseudorapidity ($\eta$) distributions. This
reduces the effective sample size by about 40\%, but makes the pion and muon
appear almost symmetric to the tracking system. The asymmetry from the pion PID
requirements is measured using a sample of unbiased
$\Dstarm\to\Dzb(\to\Kp\pim)\pim$ decays, weighted to match the \pt and $\eta$
distributions of the hard pions in the signal decays. The asymmetry from the
muon PID and trigger requirements is measured using a low-background sample of
$\jpsi\to\mup\mun$ decays with both muons reconstructed in the tracking system
and with at least one muon without trigger and muon identification
requirements. The \jpsi candidates are weighted such that the muons have the
same \pt and $\eta$ distributions as those in the signal decays.

For the $\Kp\pim$ component, the detection asymmetry is determined using prompt
\Dm decays into $\Kp\pim\pim$ and $\Kz(\to\pip\pim)\pim$ final
states~\cite{LHCb-PAPER-2014-013}. This method assumes no direct \CP violation
in these two decay modes. The candidates in the calibration samples have the
same PID requirements as those in the signal samples. The calibration samples
are weighted such that the kinematic distributions of the particles agree with
those of the kaon and soft pion in the signal samples. A small correction is
applied to account for the \Kz detection and \CP
asymmetry~\cite{LHCb-PAPER-2014-013}. The average $\Kp\pim$ detection asymmetry
is dominated by the difference in the nuclear interaction cross sections of \Kp
and \Km mesons of approximately $1\%$.

The values of \asld and \AP are determined from a two-dimensional maximum
likelihood fit to the binned distributions of \Bd decay time and charm meson
mass, simultaneously for both $f$ and \fbar final states. The fit model consists
of components for signal, background from \Bu decays to the same final state,
and combinatorial background in the \D mass distributions. The \Bu background
comes from semileptonic \Bu decays into \DstDMuNu and at least one other charged
particle. As this background is difficult to distinguish from \Bd signal decays,
fractions of this fit component are obtained from simulation and fixed in the
fit to $(12.7 \pm 2.2)\%$ for the \DMu sample and $(8.8 \pm 2.2)\%$ for the
\DstMu sample. The uncertainties are dominated by the knowledge of the branching
fractions.

\begin{figure} 
\begin{center} 
  \includegraphics[width=\figOneWidth\textwidth]{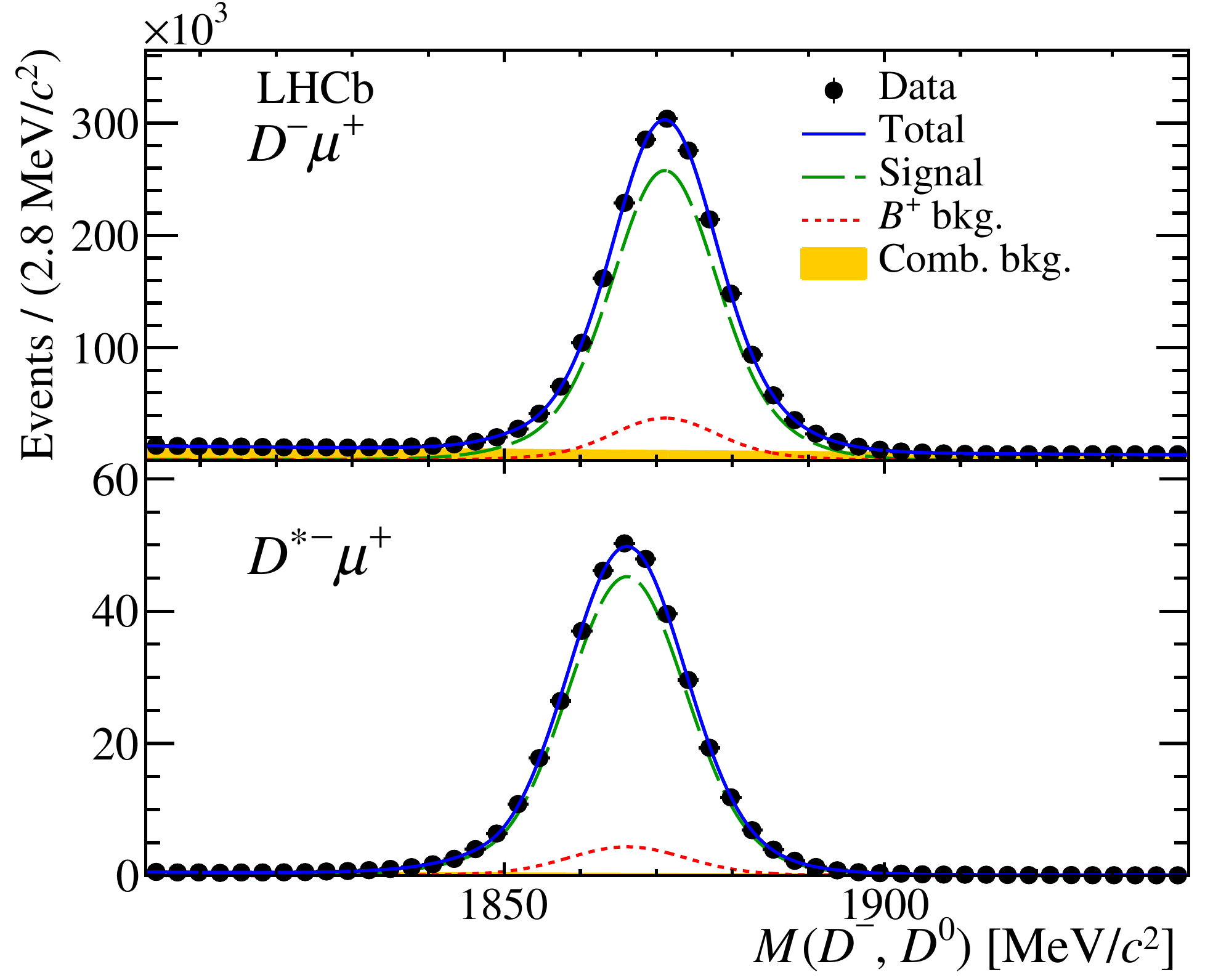}
  \caption{Mass distributions after weighting of (top) \Dm candidates in the
    \DMu sample and of (bottom) \Dstarm candidates in the \DstMu sample, with
    fit results overlaid.}
\label{fig:Dmass_fits} 
\end{center} 
\end{figure}

The mass distributions for \Dm and \Dz candidates are shown in
Fig.~\ref{fig:Dmass_fits}. To describe the mass distributions, the signal and
\Bu background are modelled by a sum of two Gaussian functions with a power-law
tail, and the combinatorial background by an exponential function.

To describe the time distributions, the signal is modelled by the decay rates of
Eq.~(\ref{eq:decayrates}).  The \Bd decay time is estimated from the \Bd flight
distance $L$, the \DstDMu momentum $p$ and the known \Bd mass
$m_{B}$~\cite{PDG2014} as $t = \mean{k} m_{B} L/p$, where $\mean{k}$ represents
a statistical correction accounting for the momentum of the missing particles in
the final state. The value of $\mean{k}$ is determined from simulation as the
average ratio between the reconstructed and true momenta of the \Bd meson, $k
\equiv p_{\rm rec}/p_{\rm true}$. The value of $\mean{k}$ depends on the \DstDMu
mass and is empirically parameterised by a second-order polynomial. This
parameterisation is used to correct the \Bd decay time. After this mass
correction, the $k/\kern -0.25em\mean{k}$ distribution has an RMS of 0.14. The
decay time distribution in the fit is described as a convolution of the decay
rates with the $k/\kern -0.25em\mean{k}$ distribution.

The efficiency as a function of the estimated decay time varies due to the IP
requirements and track reconstruction effects. This is accounted for by
multiplying the convoluted decay rates with an empirical acceptance function of
the form $( 1 - e^{-(t-t_0)/\alpha}) (1 - \beta t)$, where $t_0$ and $\alpha$
describe the effect of the IP requirements, and $\beta$ describes the track
reconstruction effect. Since $\beta$ is fully correlated with the \Bd lifetime,
the latter is fixed to the known value~\cite{PDG2014}, while $\beta$ is allowed
to vary in the fit.

The decay-time model for the \Bu background is similar to that of the signal,
except that \Bu mesons do not mix.  As the momentum spectra of the \Bd and \Bu
decay products are nearly identical, the detection asymmetry is the same as that
of the signal. The \Bu production asymmetry is taken as $(-0.6 \pm 0.6)\%$ from
the observed asymmetry in $\Bu\to\jpsi\Kp$ decays~\cite{LHCb-PAPER-2014-032}
after correcting for the kaon detection and measured \CP
asymmetries~\cite{PDG2014}.

The combinatorial background in the \D meson mass is dominated by other decays
of charm hadrons produced in \bquark-hadron decays. Hence, the decay-time model
is the same as for the signal, but setting \asld to zero.  The corresponding
values for \AP and \AD are allowed to vary in the fit.

In summary, the parameters related to the \Bu background, the detection
asymmetry, \dmd, \Gd, $t_0$, and the power-law tail in the mass distributions
are fixed in the fit; all other parameters are allowed to vary. The fit is done
in the decay-time interval $[1,15]\ps$. The effective \Bz signal yield after
weighting is 1.8 million in the \DMu sample and 0.33 million in the \DstMu
sample.

\begin{figure}[p]
  \begin{center}
  \includegraphics[width=\figTwoWidth\textwidth]{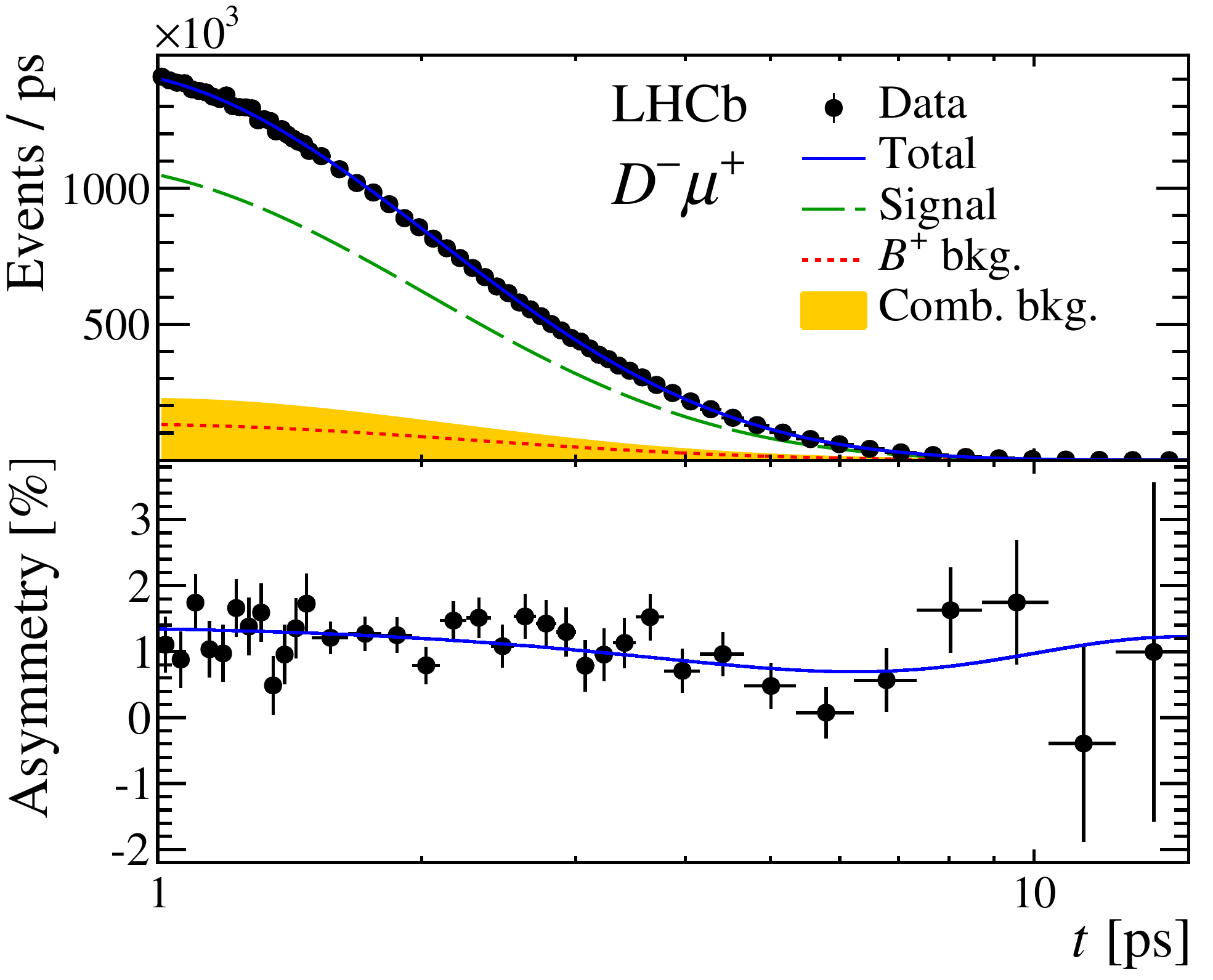}
  \includegraphics[width=\figTwoWidth\textwidth]{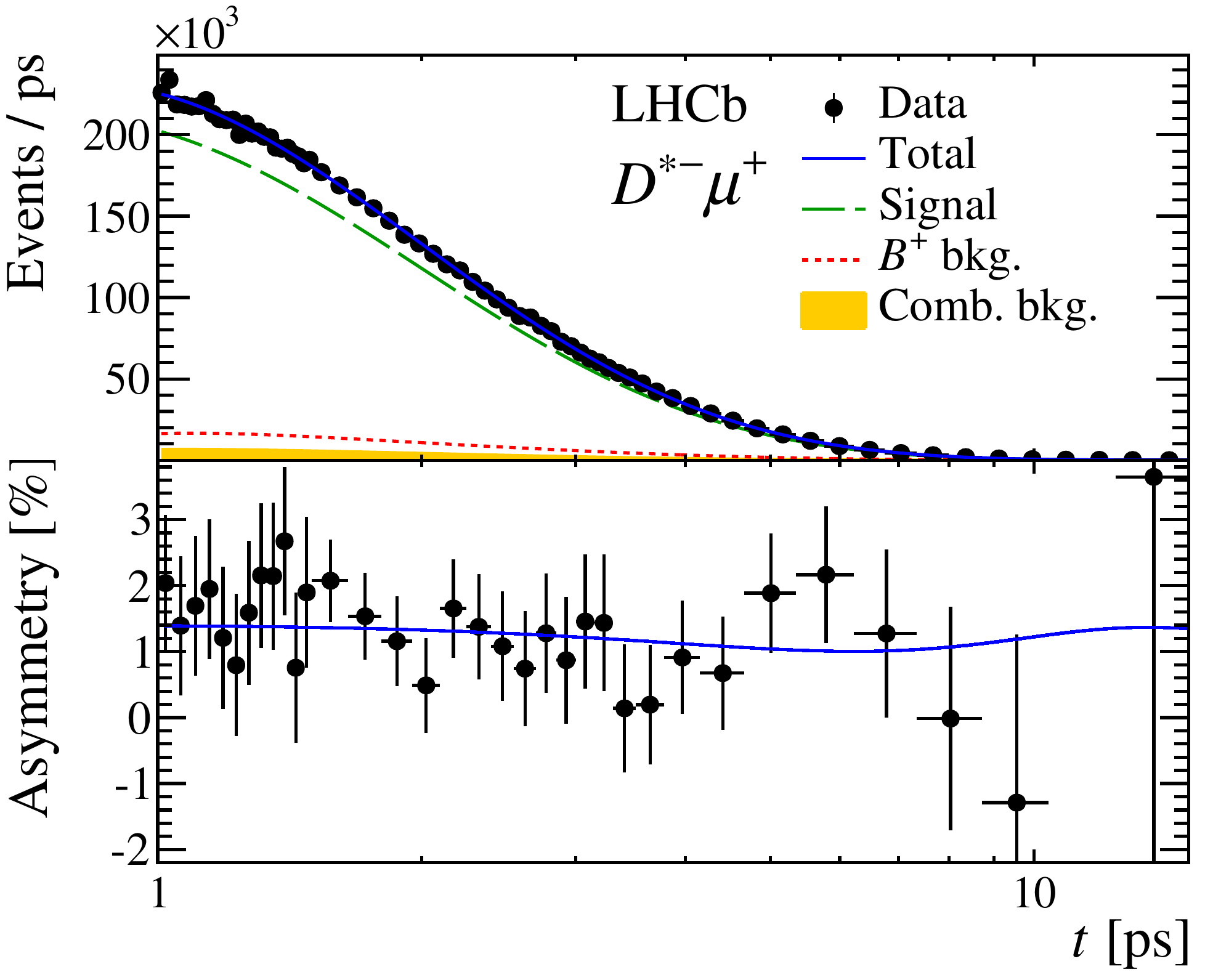}
  \caption{Decay rate and charge asymmetry after weighting versus decay time for
    (top) the \DMu sample and (bottom) the \DstMu sample. The data from the two
    run periods and magnet polarities are combined and the fit results are
    overlaid. The number of bins in the asymmetry plots is reduced for
    clarity. The visible asymmetry in these plots can be fully attributed to the
    non-zero detection and production asymmetries (not to \asld), as explained
    in the text.}
  \label{fig:charge_asymmetry}
  \end{center}
\end{figure}

Separate fits are done for the two magnet polarities, the 2011 and 2012
data-taking periods, and for the \DMu and \DstMu samples.  To reduce the bias
from any possible, unaccounted detection asymmetry, the arithmetic average of
the measured values for the two magnet polarities is taken. The resulting \asld
values for the 2011 and 2012 run periods are combined with a weighted
average. This gives $\asld = (\asldValDMu \pm \asldStatDMu)\%$ for the \DMu
sample and $\asld = (\asldValDstMu \pm \asldStatDstMu)\%$ for the \DstMu sample,
where the uncertainties are only statistical. The production asymmetries are not
averaged between the run periods as they may depend on the $pp$ centre-of-mass
energy. The decay rates and charge asymmetries as functions of the corrected
decay time are shown in Fig.~\ref{fig:charge_asymmetry}. The weighted averages
from the \DMu and \DstMu samples are used to determine the final results.  The
separate fits give compatible results for \asld and \AP. The largest difference
is seen in the 2011 data for opposite magnet polarities, where \asld differs by
about two standard deviations. This is present in both decay modes and may arise
from a statistical fluctuation of the detection asymmetry, which is highly
correlated between the two decay samples. This difference is not seen in the
larger 2012 data set.

\afterpage{\clearpage}

The systematic uncertainties are listed in Table~\ref{tab:sys_overall}.  The
largest contribution comes from the detection asymmetry, where the dominant
uncertainty is due to the limited size of the calibration samples. Additional
uncertainties are assigned to account for background in the calibration samples,
and the corresponding weighting procedures. The systematic effect from any
residual tracking asymmetry is estimated using $\jpsi\to\mup\mun$
decays~\cite{LHCb-DP-2013-002}.  The uncertainty from a possible pion
nuclear-interaction charge asymmetry is estimated to be $0.035\%$, using a
parameterisation~\cite{PDG2014} of the measured cross sections of pions on
deuterium~\cite{Galbraith, *Carroll74, *Carroll76, *Carroll79}, and the \lhcb
detector simulation.

The second largest contribution to the systematic uncertainty comes from the
knowledge of the \Bu background, and is dominated by the \Bu production
asymmetry. Uncertainties arising from the \Bu fraction, decay-time model, and
acceptance are also taken into account. Other \bquark-hadron backgrounds are
expected from semileptonic \Lb and \Bs decays and from hadronic \B decays. The
fraction of background from $\Lb\to\DstDp\mun\neumb X_n$ decays, where $X_n$
represents any neutral baryonic state, is estimated to be roughly 2\% using the
ratio of \Lb to \Bd production cross sections~\cite{LHCb-PAPER-2011-018},
simulated efficiencies, and the branching ratio of $\Lb\to\Dz\Pp\pim$ relative
to that of $\Lb\to\Lc\pim$ decays~\cite{LHCb-PAPER-2013-056}. The \Lb production
asymmetry is estimated to be $(-0.9\pm1.5)\%$, determined from the raw asymmetry
observed in $\Lb\to\jpsi\Pp\Km$~\cite{LHCb-PAPER-2014-020} and subtracting kaon
and proton detection asymmetries. The uncertainty on the \Lb production
asymmetry results in a systematic uncertainty on \asld of $0.07\%$.  The
systematic effect from an estimated $2\%$ contribution from \Bs decays is small,
since the production asymmetry vanishes due to the fast \Bs oscillations.
Hadronic decays $\B\to\DstDm\PD\PX$, where the $\PD$ meson decays
semileptonically to produce a muon, have a different $k$-factor distribution
compared to the signal. Simulation shows that these decays correspond to
approximately $1\%$ of the data and their effect is negligible. The systematic
effect from the combinatorial background in the \D mass distributions is
assessed by varying the mass model in the fit.

The uncertainty on the shape of the $k$-factor distributions comes from
uncertainties in the semileptonic branching fractions of \Bd mesons to
higher-mass \D resonances. Such decays are considered as signal, but have
slightly different $k$-factor distributions. In the \DMu sample about half of
the \Dm candidates originate from higher-mass \D resonances. The uncertainties
on these fractions are about 2\%. The systematic effect on \asld and \AP is
determined by varying the fractions by 10\% to account for possible unknown
intermediate states. The effect of a dependence of the $k$-factor with \Bd decay
time is small, and the effect on the difference in the \B momentum distributions
between data and simulation, evaluated using $\Bu\to\jpsi\Kp$ decays, is
negligible.

Systematic effects due to imperfect modelling of the decay time are tested by
varying the acceptance function and extending the fit region down to
$0.4\ps$. The effect from varying \dmd within its uncertainty~\cite{PDG2014} is
taken into account. Effects associated with variations in \Bd decay-time binning
are negligible.

\begin{table}
  \begin{center}
  \caption{Systematic uncertainties (in \%) on \asld and \AP for 7 and 8\tev
    $pp$ centre-of-mass energies. Entries marked with -- are found to be
    negligible.}
    \label{tab:sys_overall}
    \vspace{0.1cm}
    \begin{tabular}{l c c c}
      \hline \hline
\ifthenelse{\boolean{wordcount}}{ }{
 Source of uncertainty       &  ~~\asld~~ & $\AP(7\tev)$ & $\AP(8\tev)$ \\
 \hline
 Detection asymmetry         &  0.26      &  0.20        &  0.14        \\
 \Bu background              &  0.13      &  0.06        &  0.06        \\
 \Lb background              &  0.07      &  0.03        &  0.03        \\ 
 \Bs background              &  0.03      &  0.01        &  0.01        \\ 
 Combinatorial \D background &  0.03      &   --         &   --         \\
 $k$-factor distribution     &  0.03      &  0.01        &  0.01        \\
 Decay-time acceptance       &  0.03      &  0.07        &  0.07        \\ 
 Knowledge of \dmd           &  0.02      &  0.01        &  0.01        \\
 \hline
 Quadratic sum               & \asldSyst  & \ApSystSeven & \ApSystEight \\
 \hline \hline
}
    \end{tabular}
  \end{center}
\end{table}

The \Bdb--\Bd production asymmetries for the two centre-of-mass energies are
$\AP(7\tev) = (\ApValSeven \pm \ApStatSeven \pm \ApSystSeven)\%$ and $\AP(8\tev)
= (\ApValEight \pm \ApStatEight \pm \ApSystEight)\%$, where the first
uncertainty is statistical and the second systematic.  These asymmetries refer
to \Bd mesons in the ranges $2<\pt<30\gevc$ and $2.0<\eta<4.8$, without
correcting for \pt- and $\eta$-dependent reconstruction efficiencies. The
production asymmetry at 7\tev is compatible with previous
results~\cite{LHCb-PAPER-2014-042} and with the production asymmetry at 8\tev.
The determination of the \CP asymmetry in semileptonic \Bd decays is
\begin{align*}
   \asld      &= (\asldVal \pm \asldStat \pm \asldSyst)\% \ ,
\end{align*}
which is the most precise measurement to date and compatible with the SM
prediction and earlier measurements~\cite{HFAG}.

\section*{Acknowledgements}

\noindent 
We express our gratitude to our colleagues in the CERN
accelerator departments for the excellent performance of the LHC. We
thank the technical and administrative staff at the LHCb
institutes. We acknowledge support from CERN and from the national
agencies: CAPES, CNPq, FAPERJ and FINEP (Brazil); NSFC (China);
CNRS/IN2P3 (France); BMBF, DFG, HGF and MPG (Germany); SFI (Ireland); INFN (Italy); 
FOM and NWO (The Netherlands); MNiSW and NCN (Poland); MEN/IFA (Romania); 
MinES and FANO (Russia); MinECo (Spain); SNSF and SER (Switzerland); 
NASU (Ukraine); STFC (United Kingdom); NSF (USA).
The Tier1 computing centres are supported by IN2P3 (France), KIT and BMBF 
(Germany), INFN (Italy), NWO and SURF (The Netherlands), PIC (Spain), GridPP 
(United Kingdom).
We are indebted to the communities behind the multiple open 
source software packages on which we depend. We are also thankful for the 
computing resources and the access to software R\&D tools provided by Yandex LLC (Russia).
Individual groups or members have received support from 
EPLANET, Marie Sk\l{}odowska-Curie Actions and ERC (European Union), 
Conseil g\'{e}n\'{e}ral de Haute-Savoie, Labex ENIGMASS and OCEVU, 
R\'{e}gion Auvergne (France), RFBR (Russia), XuntaGal and GENCAT (Spain), Royal Society and Royal
Commission for the Exhibition of 1851 (United Kingdom).

\addcontentsline{toc}{section}{References}
\setboolean{inbibliography}{true}
\bibliographystyle{LHCb}
\bibliography{main,LHCb-PAPER,LHCb-DP}



\newpage


\centerline{\large\bf LHCb collaboration}
\begin{flushleft}
\small
R.~Aaij$^{41}$, 
B.~Adeva$^{37}$, 
M.~Adinolfi$^{46}$, 
A.~Affolder$^{52}$, 
Z.~Ajaltouni$^{5}$, 
S.~Akar$^{6}$, 
J.~Albrecht$^{9}$, 
F.~Alessio$^{38}$, 
M.~Alexander$^{51}$, 
S.~Ali$^{41}$
% ...

{\footnotesize \it
% ...
$ ^{5}$Clermont Universit\'{e}, Universit\'{e} Blaise Pascal, CNRS/IN2P3, LPC, Clermont-Ferrand, France\\
$ ^{6}$CPPM, Aix-Marseille Universit\'{e}, CNRS/IN2P3, Marseille, France\\
% ...
$ ^{9}$Fakult\"{a}t Physik, Technische Universit\"{a}t Dortmund, Dortmund, Germany\\
% ...
$ ^{37}$Universidad de Santiago de Compostela, Santiago de Compostela, Spain\\
$ ^{38}$European Organization for Nuclear Research (CERN), Geneva, Switzerland\\
% ...
$ ^{41}$Nikhef National Institute for Subatomic Physics, Amsterdam, The Netherlands\\
% ...
$ ^{46}$H.H. Wills Physics Laboratory, University of Bristol, Bristol, United Kingdom\\
% ...
$ ^{51}$School of Physics and Astronomy, University of Glasgow, Glasgow, United Kingdom\\
$ ^{52}$Oliver Lodge Laboratory, University of Liverpool, Liverpool, United Kingdom\\
% ...
}
\end{flushleft}

\end{document}